\documentclass[journal=apchd5,manuscript=article,floatfix,10pt,superscriptaddress]{achemso}

\usepackage{bm}
\usepackage{graphicx}
\usepackage{subfigure}
\usepackage{epstopdf}
\usepackage{siunitx}
\usepackage{braket}
\usepackage{tabularx}
\usepackage{blindtext}
\usepackage{amsmath}
\usepackage{natbib}
\usepackage{float}
\usepackage{color}

\def\bem#1{\begin{mathletters}\label{#1}}
	\def\eml{\end{mathletters}}

\def\4#1{{\boldsymbol{#1}}}
\def\8#1{{\widetilde{#1}}}

\title {High-Sensitivity, High-Resolution Detection of Reactive Oxygen-Species Concentration Using NV Centers}

\author{Yoav Ninio}
\affiliation{Dept. of Applied Physics, The Hebrew University of Jerusalem, Safra Campus, Givat Ram, Jerusalem 91904, Israel}
\alsoaffiliation{Center for Nanoscience and Nanotechnology, The Hebrew University of Jerusalem, Safra Campus, Givat Ram, Jerusalem 91904, Israel}

\author{Nir Waiskopf}
\affiliation{Institute of Chemistry, The Hebrew University of Jerusalem, Safra Campus, Givat Ram, Jerusalem 91904, Israel}
\alsoaffiliation{Center for Nanoscience and Nanotechnology, The Hebrew University of Jerusalem, Safra Campus, Givat Ram, Jerusalem 91904, Israel}

\author{Idan Meirzada}
\affiliation{The Racah Institute of Physics, The Hebrew University of Jerusalem, Safra Campus, Givat Ram, Jerusalem 91904, Israel}
\alsoaffiliation{Center for Nanoscience and Nanotechnology, The Hebrew University of Jerusalem, Safra Campus, Givat Ram, Jerusalem 91904, Israel}

\author{Yoav Romach}
\affiliation{The Racah Institute of Physics, The Hebrew University of Jerusalem, Safra Campus, Givat Ram, Jerusalem 91904, Israel}
\alsoaffiliation{Center for Nanoscience and Nanotechnology, The Hebrew University of Jerusalem, Safra Campus, Givat Ram, Jerusalem 91904, Israel}

\author{Galya Haim}
\affiliation{Dept. of Applied Physics, The Hebrew University of Jerusalem, Safra Campus, Givat Ram, Jerusalem 91904, Israel}
\alsoaffiliation{Center for Nanoscience and Nanotechnology, The Hebrew University of Jerusalem, Safra Campus, Givat Ram, Jerusalem 91904, Israel}

\author{Shira Yochelis}
\affiliation{Dept. of Applied Physics, The Hebrew University of Jerusalem, Safra Campus, Givat Ram, Jerusalem 91904, Israel}
\alsoaffiliation{Center for Nanoscience and Nanotechnology, The Hebrew University of Jerusalem, Safra Campus, Givat Ram, Jerusalem 91904, Israel}

\author{Uri Banin}
\affiliation{Institute of Chemistry, The Hebrew University of Jerusalem, Safra Campus, Givat Ram, Jerusalem 91904, Israel}
\alsoaffiliation{Center for Nanoscience and Nanotechnology, The Hebrew University of Jerusalem, Safra Campus, Givat Ram, Jerusalem 91904, Israel}

\author{Nir Bar-Gill}
\email{bargill@phys.huji.ac.il}

\affiliation{Dept. of Applied Physics, The Hebrew University of Jerusalem, Safra Campus, Givat Ram, Jerusalem 91904, Israel}
\alsoaffiliation{The Racah Institute of Physics, The Hebrew University of Jerusalem, Safra Campus, Givat Ram, Jerusalem 91904, Israel}

\keywords{Nitrogen-Vacancy , Fluorescence, Probe intracellular signals, Reduction–oxidation}

\begin{document}

	\begin{abstract}
		Nitrogen-vacancy (NV) color centers in diamond have been demonstrated as useful magnetic sensors, in particular for measuring spin fluctuations, achieving high sensitivity and spatial resolution. These abilities can be used to explore various biological and chemical processes, catalyzed by Reactive Oxygen Species (ROS). Here we demonstrate a novel approach to measure and quantify Hydroxyl radicals with high spatial resolution, using the fluorescence difference between NV charged states. According to the results, the achieved NV sensitivity is $11 \pm 4 \frac{nM}{\sqrt Hz}$, realized in-situ without spin labels and localized to a volume of $\sim 10$ picoliter.
	\end{abstract}

\section{Introduction}
Reactive Oxygen Species (ROS) are chemically reactive molecules containing oxygen. ROS are quite ubiquitous, participating in various catalytic processes \cite{Banin2018,Xu2020,Zhou2020}, as well as in biochemical reactions and cellular signaling, including pathological situations such as cancer, cardiovascular pathologies and autoimmune diseases \cite{wu_production_2011,Vaccaro2020,Rowe2020,Sies2020}. Due to their importance, the ability to characterize ROS with high sensitivity and spatial resolution could provide insights into the local dynamics of various processes in Chemistry and Biology. 

The presence of unpaired spins in ROS generates magnetic noise which can be detected using proximal magnetic sensors. However, due to their highly reactive nature, their lifetime is relatively short, usually on the order of a few $\mu$s \cite{attri_generation_2015}, making such measurements difficult. The current state-of-the-art methods for detecting ROS, such as spectrophotometry and chemiluminescence, rely on processes in which they create stable labels \cite{pavelescu_reactive_2015}. Using these methods, it is nowadays possible to measure radical concentrations down to $5.3\times 10^{-13}\frac{M}{\sqrt Hz}$ \cite{lankone_uvvis_2020}. Despite the high sensitivity of the aforementioned methods, they are designed for relatively large volumes, and thus are limited in their ability to provide spatial information on the radical concentration. 

\begin{figure}[tbh]
\begin{center}
{\includegraphics[width=4.5cm ,trim=2cm 0cm 2cm 0cm ]{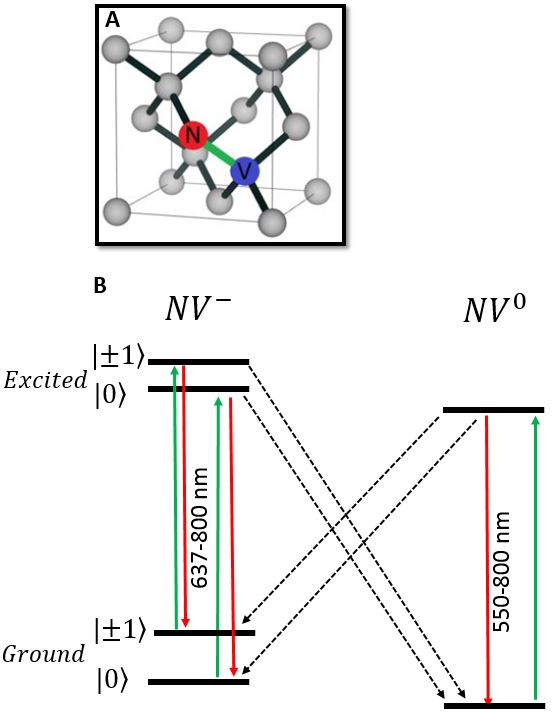}}
\end{center}
\caption{Energy level and crystallographic structure of NV center. Figure 1.A shows Nitrogen-Vacancy in a diamond lattice\cite{momma_vesta_2011}. Figure 1.B represent Energy level diagram. Green excitation
is depicted with green arrows, red fluorescence is depicted with downward red arrows.Transition between NV$^-$ excited state and NV$^0$ ground state is depicted with dashed arrows \cite{meirzada_enhanced_2019}.  } 
\label{fig:NV}
\end{figure}

In recent years, nitrogen-vacancy (NV) centers in diamond have emerged as useful magnetic sensors, due to their unique spin and optical properties \cite{Doherty2013,Liu2016,Moscariello2019,Han2019}. The NV center is a color defect in the diamond lattice, comprising the combination of a carbon vacancy and an adjacent nitrogen atom. As illustrated in Figure \ref{fig:NV}, the NV center can be found in two charged states: neutral NV$^0$ and negative NV$^-$, which have different emission spectra \cite{rondin_surface-induced_2010}. 
NV-based magnetic noise measurements have been demonstrated in various contexts, essentially performing relaxometry, i.e. monitoring the effect of magnetic field fluctuations on the NV relaxation time \cite{kaufmann_detection_2013,steinert_magnetic_2013,McCullian2020}. Previous work focused on characterizing spin concentrations through relaxometry was focused mostly on Gadolinium$^{+3}$, a common MRI contrast agent with spin 7/2, which produces characteristic magnetic fluctuations that can reduce the $T_1$ relaxation time by more than 90\% \cite{steinert_magnetic_2013}.

\begin{figure*}[tbh]
\begin{center}
{\includegraphics[width=6cm ,trim=8cm 0cm 8cm 0cm ]{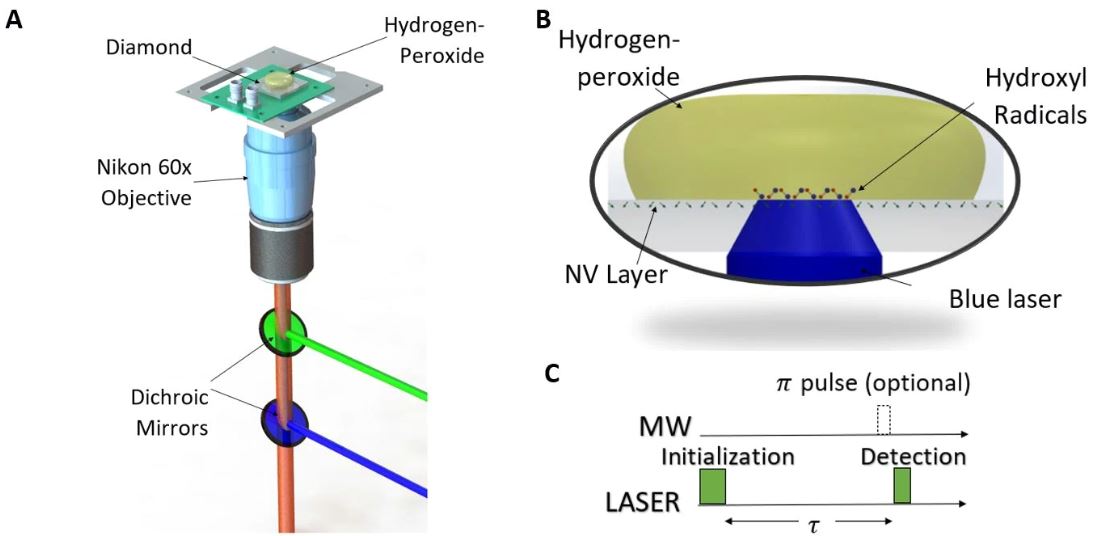}}
\end{center}
\caption{(A) Schematic representation of the confocal setup. (B) Side view of the diamond covered with Hydrogen Peroxide. 2.5 $\mu L$ Hydrogen peroxide was deposited on the top of the diamond. Blue laser illuminated the Hydrogen peroxide through  the diamond to generate Hydroxyl radicals, adjacent to the NV layer. (C) T1 sequence. The NV's were first polarized to the $|0 \rangle$ ground state, and a fluorescence measurement was conducted after a delay of $\tau$ . For improved contrast, in some instances the same sequence was conducted with a microwave (MW) $\pi$ pulse (flips the spin to $|\pm 1 \rangle$ ) before detection. }
\label{fig:setup}
\end{figure*}

In this work we suggest and demonstrate a novel approach for the detection and quantification of radical concentrations with high spatial resolution. Our method is based on NV$^-$ and NV$^0$ fluorescence measurements, wherein the presence of radicals provides a pathway for electron relaxation, thus decreasing the NV$^-$ fluorescence rate, while increasing the NV$^0$ fluorescence rate.

The experiments were performed using a home-built NV microscope, depicted schematically in Figure \ref{fig:setup}a. The sample (Gd or Hydrogen-peroxide) was placed on the diamond surface (Figure \ref{fig:setup}b). Basic relaxometry was performed using a standard longitudinal relaxation measurement sequence, $T_1$, shown in Figure \ref{fig:setup}c.

\section{Results and Discussion}
\begin{figure}[tbh]
\begin{center}
{\includegraphics[width=8cm ,trim=6cm 0cm 6cm 0cm ]{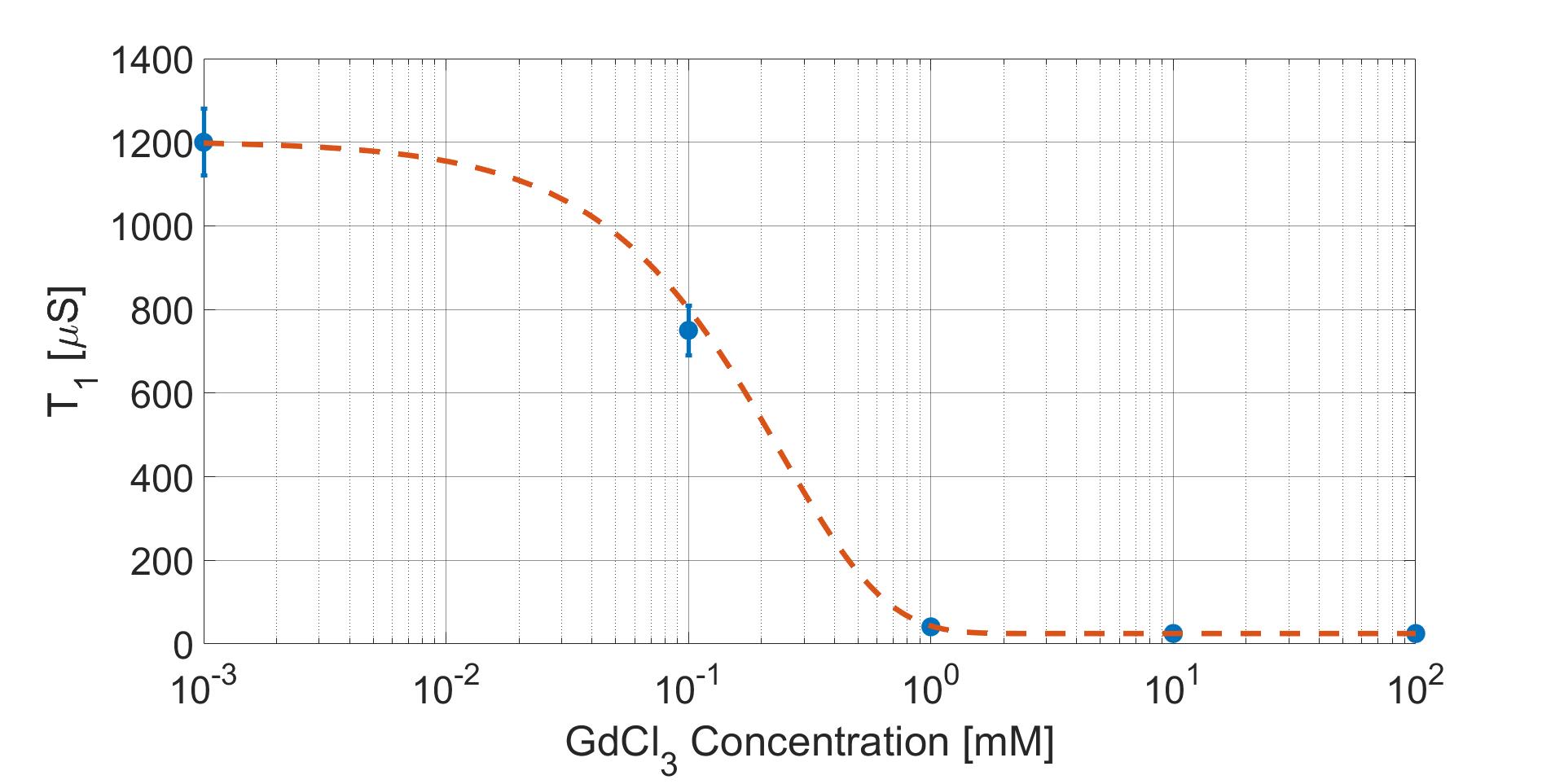}}
\end{center}
\caption{Relaxation time ($T_1$) in different GdCl$_3$ concentrations. Relaxation time defined as the decay constant at $T_1$ experiment.
Each blue dot represents relaxation time in different GdCl$_3$ concentration. Dashed curve is an exponential fit for the change in relaxation time as a function of GdCl$_3$ concentration.} 
\label{fig:Gd}
\end{figure}

We initially implemented in our system standard relaxometry protocols, studying the change in the longitudinal $T_1$ relaxation time of the NVs in the presence of magnetic noise from Gadolinium (III) chloride (GdCl$_3$) as a function of concentration.
The Gadolinium(III) chloride was mixed with Agar, forming a solid matrix with uniform GdCl$_3$ concentration. Subsequently, the mixture was deposited on the top of the diamond. The results are summarized in Figure \ref{fig:Gd}, depicting the dependence of the measured $T_1$ relaxation time on the Gd concentration. Comparing to previously published work \cite{kaufmann_detection_2013,steinert_magnetic_2013,McCullian2020} we find that our sensitivity, which is limited to a concentration of approximately $100 \mu$M, slightly exceeds the state-of-the-art and indicates an appropriate experimental setup and diamond sample. 

This approach, however, is hindered by the measurement timescale (corresponding to the $T_1$ relaxation time, on the order of ms), which is long compared to the ROS lifetime. This could potentially lead to variations in the ROS concentrations during the measurement time. In addition, these experiments have limited Signal to Noise ratio (SNR). 

In the following we present a faster scheme based on NV fluorescence changes resulting from quenching induced by the radicals.

\begin{figure}[tbh]
\begin{center}
{\includegraphics[width=0.55 \linewidth]{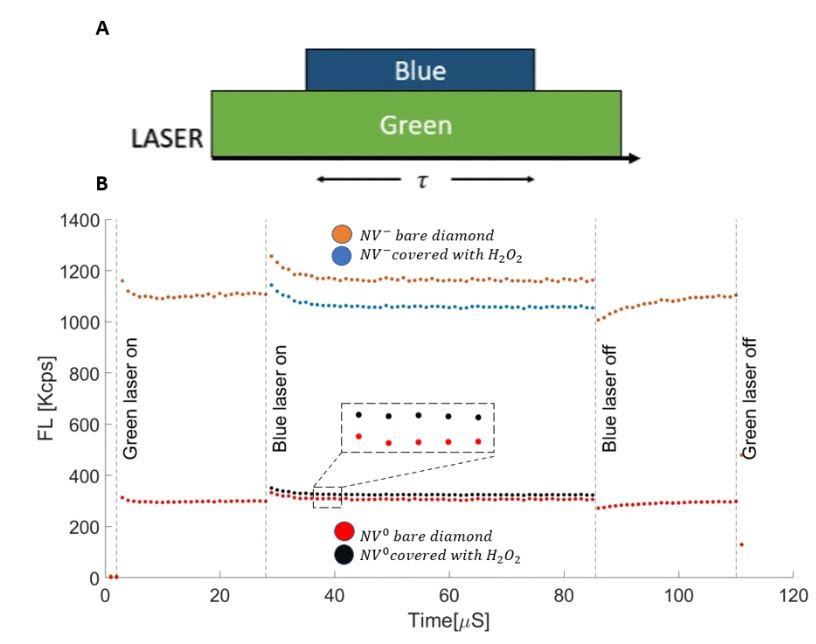}}
\end{center}
\caption{(a) Fluorescence sequence. The blue laser was turned on for a period $\tau$, while the green laser was on constantly. The fluorescence was recorded during the entire experiment. (b) Comparison between Fluorescence sequences. Orange and blue points represents experiments using $NV^-$ fluorescence filter. Orange points represent fluorescence with bare diamond while blue points presents fluorescence in the presence of Hydrogen-Peroxide. Red and black points represent experiments using an $NV^0$ fluorescence filter. Black points represent fluorescence with bare diamond while red points represent fluorescence with Hydrogen-Peroxide. The blue laser power during all experiments was 217$\mu$W.} 
\label{fig:flSeq}
\end{figure}

Figure \ref{fig:flSeq}a presents the  experimental protocol for fluorescence measurements. In these experiments, we  generated radicals by illuminating hydrogen-peroxide with a blue laser (405 nm). This induces a photo-chemical process which decomposes the H$_2$O$_2$ molecule to form two Hydroxyl radicals. This allows us to controllably create radicals at a desired rate, corresponding to the intensity of the blue light. Specifically, we can measure in-situ the fluorescence change induced by the radicals (FL), compared to the fluorescence of the baseline value, obtained with the same sequence for bare diamond.
The change in fluorescence was calculated as follows :
\begin{equation}
\textrm{Change in FL}= \frac{\textrm{Average FL with blue laser }}{\textrm{Average FL without blue laser }}
\end{equation} 
Figure \ref{fig:flSeq}b depicts measured fluorescence during the sequence as a function of time. We performed these measurements, extracting the change in fluorescence of both NV$^-$ and NV$^0$ with ROS and for bare diamond. We can see that as the blue laser is turned on and off, a transient of $<10 \mu$s precedes the steady-state fluorescence value.

We note that the blue laser (405 nm), used to generate radicals, causes an increase of the fluorescence intensity in both NV$^-$ and NV$^0$ spectra (Fig. \ref{fig:flSeq}b). This is consistent with the excitation of the NV centers by both green and blue lasers. Therefore the added excitation leads to a higher population in the NV excited state which can decay radiatively. 

\begin{figure}[tbh]
\begin{center}
{\includegraphics[width=0.5 \linewidth]{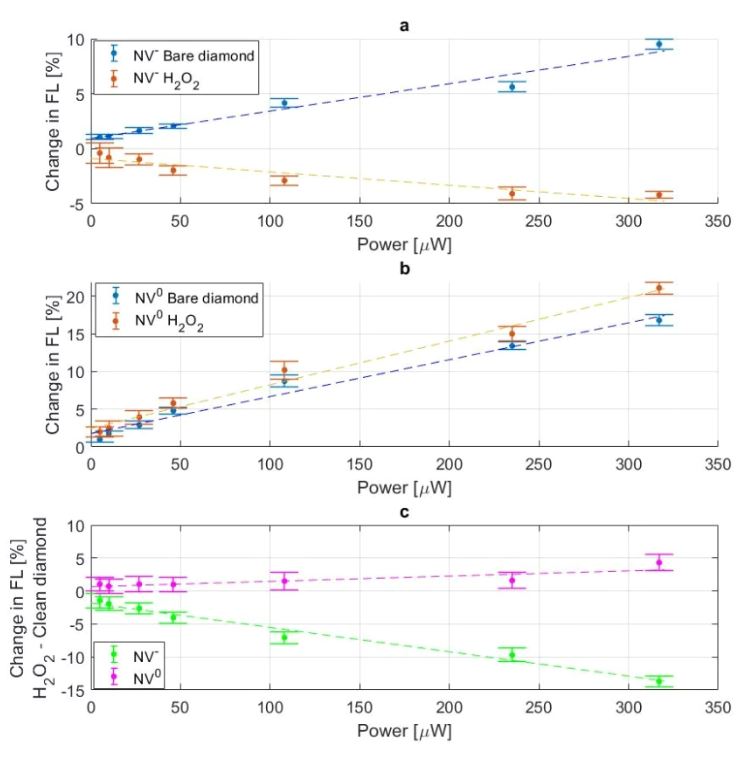}}
\end{center}
\caption{Comparison between the change in fluorescence with and without hydrogen peroxide. The average fluorescence during blue laser illumination was divided by the average fluorescence without the blue laser. (a) The change in fluorescence of the $NV^-$ spectrum vs. blue laser power. (b) The change in fluorescence of the  $NV^0$ spectrum vs. blue laser power. (c) Normalized change in fluorescence, subtracting the results with a bare diamond to the results with hydrogen-peroxide, of the $NV^-$ spectrum (blue) and the $NV^0$ spectrum (red) as a function of blue laser power. }
\label{fig:results}
\end{figure}

Figure \ref{fig:results} summarizes the main results of the paper - showing the steady-state change in normalized fluorescence (FL) as a function of radical concentration (controlled by the intensity of the applied blue light). In this set of experiments $2.5 \mu$L of Hydrogen-Peroxide ($H_2 O_2$, concentration=30\%) covered the top of the diamond (Fig. \ref{fig:setup}b). Upon increasing the blue laser power, a  drop in $NV^-$ fluorescence is seen (Figure \ref{fig:results}a), accompanied by an increase in the $NV^0$ fluorescence (Figure \ref{fig:results}b).
Figure \ref{fig:results}c presents the normalized fluorescence change with respect to the bare diamond results as a function of blue light intensity, clearly showing the increased NV$^0$ fluorescence concurrently with the drop in NV$^-$ fluorescence. Calibrating the blue illumination intensity to ROS concentration, this fluorescence change can be translated to radical concentration sensitivity. 

As mentioned above, the blue laser decomposes the H$_2 $O$_2$ molecule and forms Hydroxyl radicals which are highly reactive. Thus, a redox reaction occur between the NV (undergoes oxidation) and the radicals. This process is consistent with the rise in NV$^0$ fluorescence at the expense of NV$^-$ fluorescence (Fig. \ref{fig:results}). To translate the blue laser intensity to Hydroxyl concentration, we conducted a calibrated Beer-Lambert absorption experiment (see appendix B). These measurements indicate that at threshold illumination (5$ \mu$W) the steady-state Hydroxyl concentration is 16nM (see Appendix C). Given this calibration, we deduce that our NV fluorescence measurement sensitivity to radical concentration is $11 \pm 4 \frac{nM}{\sqrt Hz}$. We note that the NV system allows localized, small volume measurements, down to $\sim 10$ picoliter in our experiment (for which we can translate the molar concentration sensitivity to an effective number of molecules, which is found to be much smaller than 1, i.e. $0.03 \pm 0.01 \frac{Particles}{\sqrt Hz}$).

To summarize, we presented a novel scheme for local, high sensitivity characterization of radical concentration, based on fluorescence variations of NV centers in diamond. We demonstrate this technique with hydroxil radicals activated from Hydrogen-peroxide using blue light illumination, and find a sensitivity of $11 \pm 4 \frac{nM}{\sqrt{Hz}}$ in a sub $1 \mu m^3$ volume. We note that the NV sensing platform is all optical, and robust to environmental conditions. Thus, these results introduce a new, non-destructive tool for in-situ radical characterization with sub-micron resolution and nM sensitivity. 
This opens up future opportunities for the direct probing and spatial mapping of radical formation and dynamics. This has wide ranging implications: (i) In Material Science and Chemistry, for example 
the study of radical related battery degradation; (ii) In Biology, for example in high resolution mapping and tracking of ROS-mediated intracellular signal transduction and cellular signaling, providing valuable information on the origin, propagation and functioning of these ROS in biological systems.

\section{Acknowledgments}
Y.N. acknowledges support from the Council for Higher Education, Israel, and from the The Peter Brojde Center for Innovative Engineering and Computer Science at the Hebrew University.

N.B. acknowledges support from the European Union’s Horizon 2020 research and innovation program under grant agreements No. 714005 (ERC StG Q-DIM-SIM), No. 820374 (MetaboliQs), and No. 828946 (PATHOS), and has been supported in part by the Ministry of Science and Technology, Israel. U.B acknowledges support by the Israel Science Foundation (grant No. 1363/18 ). U.B. thanks the Alfred \& Erica Larisch memorial chair.

\bibliographystyle{MSP}
\bibliography{paper.bib.bib}

\begin{thebibliography}{28}%
\makeatletter
\providecommand \@ifxundefined [1]{%
 \@ifx{#1\undefined}
}%
\providecommand \@ifnum [1]{%
 \ifnum #1\expandafter \@firstoftwo
 \else \expandafter \@secondoftwo
 \fi
}%
\providecommand \@ifx [1]{%
 \ifx #1\expandafter \@firstoftwo
 \else \expandafter \@secondoftwo
 \fi
}%
\providecommand \natexlab [1]{#1}%
\providecommand \enquote  [1]{``#1''}%
\providecommand \bibnamefont  [1]{#1}%
\providecommand \bibfnamefont [1]{#1}%
\providecommand \citenamefont [1]{#1}%
\providecommand \href@noop [0]{\@secondoftwo}%
\providecommand \href [0]{\begingroup \@sanitize@url \@href}%
\providecommand \@href[1]{\@@startlink{#1}\@@href}%
\providecommand \@@href[1]{\endgroup#1\@@endlink}%
\providecommand \@sanitize@url [0]{\catcode `\\12\catcode `\$12\catcode
  `\&12\catcode `\#12\catcode `\^12\catcode `\_12\catcode `\%12\relax}%
\providecommand \@@startlink[1]{}%
\providecommand \@@endlink[0]{}%
\providecommand \url  [0]{\begingroup\@sanitize@url \@url }%
\providecommand \@url [1]{\endgroup\@href {#1}{\urlprefix }}%
\providecommand \urlprefix  [0]{URL }%
\providecommand \Eprint [0]{\href }%
\providecommand \doibase [0]{http://dx.doi.org/}%
\providecommand \selectlanguage [0]{\@gobble}%
\providecommand \bibinfo  [0]{\@secondoftwo}%
\providecommand \bibfield  [0]{\@secondoftwo}%
\providecommand \translation [1]{[#1]}%
\providecommand \BibitemOpen [0]{}%
\providecommand \bibitemStop [0]{}%
\providecommand \bibitemNoStop [0]{.\EOS\space}%
\providecommand \EOS [0]{\spacefactor3000\relax}%
\providecommand \BibitemShut  [1]{\csname bibitem#1\endcsname}%
\let\auto@bib@innerbib\@empty
\bibitem [{\citenamefont {Reed}\ \emph {et~al.}(2010)\citenamefont {Reed},
  \citenamefont {DiCarlo}, \citenamefont {Johnson}, \citenamefont {Sun},
  \citenamefont {Schuster}, \citenamefont {Frunzio},\ and\ \citenamefont
  {Schoelkopf}}]{reed_high-fidelity_2010}%
  \BibitemOpen
  \bibfield  {author} {\bibinfo {author} {\bibfnamefont {M.~D.}\ \bibnamefont
  {Reed}}, \bibinfo {author} {\bibfnamefont {L.}~\bibnamefont {DiCarlo}},
  \bibinfo {author} {\bibfnamefont {B.~R.}\ \bibnamefont {Johnson}}, \bibinfo
  {author} {\bibfnamefont {L.}~\bibnamefont {Sun}}, \bibinfo {author}
  {\bibfnamefont {D.~I.}\ \bibnamefont {Schuster}}, \bibinfo {author}
  {\bibfnamefont {L.}~\bibnamefont {Frunzio}}, \ and\ \bibinfo {author}
  {\bibfnamefont {R.~J.}\ \bibnamefont {Schoelkopf}},\ }\href {\doibase
  10.1103/PhysRevLett.105.173601} {\bibfield  {journal} {\bibinfo  {journal}
  {Physical Review Letters}\ }\textbf {\bibinfo {volume} {105}} (\bibinfo
  {year} {2010}),\ 10.1103/PhysRevLett.105.173601}\BibitemShut {NoStop}%
\bibitem [{\citenamefont {Myerson}\ \emph {et~al.}(2008)\citenamefont
  {Myerson}, \citenamefont {Szwer}, \citenamefont {Webster}, \citenamefont
  {Allcock}, \citenamefont {Curtis}, \citenamefont {Imreh}, \citenamefont
  {Sherman}, \citenamefont {Stacey}, \citenamefont {Steane},\ and\
  \citenamefont {Lucas}}]{myerson_high-fidelity_2008}%
  \BibitemOpen
  \bibfield  {author} {\bibinfo {author} {\bibfnamefont {A.~H.}\ \bibnamefont
  {Myerson}}, \bibinfo {author} {\bibfnamefont {D.~J.}\ \bibnamefont {Szwer}},
  \bibinfo {author} {\bibfnamefont {S.~C.}\ \bibnamefont {Webster}}, \bibinfo
  {author} {\bibfnamefont {D.~T.~C.}\ \bibnamefont {Allcock}}, \bibinfo
  {author} {\bibfnamefont {M.~J.}\ \bibnamefont {Curtis}}, \bibinfo {author}
  {\bibfnamefont {G.}~\bibnamefont {Imreh}}, \bibinfo {author} {\bibfnamefont
  {J.~A.}\ \bibnamefont {Sherman}}, \bibinfo {author} {\bibfnamefont {D.~N.}\
  \bibnamefont {Stacey}}, \bibinfo {author} {\bibfnamefont {A.~M.}\
  \bibnamefont {Steane}}, \ and\ \bibinfo {author} {\bibfnamefont {D.~M.}\
  \bibnamefont {Lucas}},\ }\href {\doibase 10.1103/PhysRevLett.100.200502}
  {\bibfield  {journal} {\bibinfo  {journal} {Physical Review Letters}\
  }\textbf {\bibinfo {volume} {100}} (\bibinfo {year} {2008}),\
  10.1103/PhysRevLett.100.200502}\BibitemShut {NoStop}%
\bibitem [{\citenamefont {Steiner}\ \emph {et~al.}(2010)\citenamefont
  {Steiner}, \citenamefont {Neumann}, \citenamefont {Beck}, \citenamefont
  {Jelezko},\ and\ \citenamefont {Wrachtrup}}]{steiner_universal_2010}%
  \BibitemOpen
  \bibfield  {author} {\bibinfo {author} {\bibfnamefont {M.}~\bibnamefont
  {Steiner}}, \bibinfo {author} {\bibfnamefont {P.}~\bibnamefont {Neumann}},
  \bibinfo {author} {\bibfnamefont {J.}~\bibnamefont {Beck}}, \bibinfo {author}
  {\bibfnamefont {F.}~\bibnamefont {Jelezko}}, \ and\ \bibinfo {author}
  {\bibfnamefont {J.}~\bibnamefont {Wrachtrup}},\ }\href {\doibase
  10.1103/PhysRevB.81.035205} {\bibfield  {journal} {\bibinfo  {journal}
  {Physical Review B}\ }\textbf {\bibinfo {volume} {81}} (\bibinfo {year}
  {2010}),\ 10.1103/PhysRevB.81.035205}\BibitemShut {NoStop}%
\bibitem [{\citenamefont {Morello}\ \emph {et~al.}(2010)\citenamefont
  {Morello}, \citenamefont {Pla}, \citenamefont {Zwanenburg}, \citenamefont
  {Chan}, \citenamefont {Tan}, \citenamefont {Huebl}, \citenamefont
  {Möttönen}, \citenamefont {Nugroho}, \citenamefont {Yang}, \citenamefont
  {van Donkelaar}, \citenamefont {Alves}, \citenamefont {Jamieson},
  \citenamefont {Escott}, \citenamefont {Hollenberg}, \citenamefont {Clark},\
  and\ \citenamefont {Dzurak}}]{morello_single-shot_2010}%
  \BibitemOpen
  \bibfield  {author} {\bibinfo {author} {\bibfnamefont {A.}~\bibnamefont
  {Morello}}, \bibinfo {author} {\bibfnamefont {J.~J.}\ \bibnamefont {Pla}},
  \bibinfo {author} {\bibfnamefont {F.~A.}\ \bibnamefont {Zwanenburg}},
  \bibinfo {author} {\bibfnamefont {K.~W.}\ \bibnamefont {Chan}}, \bibinfo
  {author} {\bibfnamefont {K.~Y.}\ \bibnamefont {Tan}}, \bibinfo {author}
  {\bibfnamefont {H.}~\bibnamefont {Huebl}}, \bibinfo {author} {\bibfnamefont
  {M.}~\bibnamefont {Möttönen}}, \bibinfo {author} {\bibfnamefont {C.~D.}\
  \bibnamefont {Nugroho}}, \bibinfo {author} {\bibfnamefont {C.}~\bibnamefont
  {Yang}}, \bibinfo {author} {\bibfnamefont {J.~A.}\ \bibnamefont {van
  Donkelaar}}, \bibinfo {author} {\bibfnamefont {A.~D.~C.}\ \bibnamefont
  {Alves}}, \bibinfo {author} {\bibfnamefont {D.~N.}\ \bibnamefont {Jamieson}},
  \bibinfo {author} {\bibfnamefont {C.~C.}\ \bibnamefont {Escott}}, \bibinfo
  {author} {\bibfnamefont {L.~C.~L.}\ \bibnamefont {Hollenberg}}, \bibinfo
  {author} {\bibfnamefont {R.~G.}\ \bibnamefont {Clark}}, \ and\ \bibinfo
  {author} {\bibfnamefont {A.~S.}\ \bibnamefont {Dzurak}},\ }\href {\doibase
  10.1038/nature09392} {\bibfield  {journal} {\bibinfo  {journal} {Nature}\
  }\textbf {\bibinfo {volume} {467}},\ \bibinfo {pages} {687} (\bibinfo {year}
  {2010})}\BibitemShut {NoStop}%
\bibitem [{\citenamefont {Fuchs}\ \emph {et~al.}(2011)\citenamefont {Fuchs},
  \citenamefont {Burkard}, \citenamefont {Klimov},\ and\ \citenamefont
  {Awschalom}}]{fuchs_quantum_2011}%
  \BibitemOpen
  \bibfield  {author} {\bibinfo {author} {\bibfnamefont {G.~D.}\ \bibnamefont
  {Fuchs}}, \bibinfo {author} {\bibfnamefont {G.}~\bibnamefont {Burkard}},
  \bibinfo {author} {\bibfnamefont {P.~V.}\ \bibnamefont {Klimov}}, \ and\
  \bibinfo {author} {\bibfnamefont {D.~D.}\ \bibnamefont {Awschalom}},\ }\href
  {\doibase 10.1038/nphys2026} {\bibfield  {journal} {\bibinfo  {journal}
  {Nature Physics}\ }\textbf {\bibinfo {volume} {7}},\ \bibinfo {pages} {789}
  (\bibinfo {year} {2011})}\BibitemShut {NoStop}%
\bibitem [{\citenamefont {Clevenson}\ \emph {et~al.}(2015)\citenamefont
  {Clevenson}, \citenamefont {Trusheim}, \citenamefont {Teale}, \citenamefont
  {Schröder}, \citenamefont {Braje},\ and\ \citenamefont
  {Englund}}]{clevenson_broadband_2015}%
  \BibitemOpen
  \bibfield  {author} {\bibinfo {author} {\bibfnamefont {H.}~\bibnamefont
  {Clevenson}}, \bibinfo {author} {\bibfnamefont {M.~E.}\ \bibnamefont
  {Trusheim}}, \bibinfo {author} {\bibfnamefont {C.}~\bibnamefont {Teale}},
  \bibinfo {author} {\bibfnamefont {T.}~\bibnamefont {Schröder}}, \bibinfo
  {author} {\bibfnamefont {D.}~\bibnamefont {Braje}}, \ and\ \bibinfo {author}
  {\bibfnamefont {D.}~\bibnamefont {Englund}},\ }\href {\doibase
  10.1038/nphys3291} {\bibfield  {journal} {\bibinfo  {journal} {Nature
  Physics}\ }\textbf {\bibinfo {volume} {11}},\ \bibinfo {pages} {393}
  (\bibinfo {year} {2015})}\BibitemShut {NoStop}%
\bibitem [{\citenamefont {Acosta}\ \emph
  {et~al.}(2010{\natexlab{a}})\citenamefont {Acosta}, \citenamefont {Bauch},
  \citenamefont {Jarmola}, \citenamefont {Zipp}, \citenamefont {Ledbetter},\
  and\ \citenamefont {Budker}}]{acosta_broadband_2010}%
  \BibitemOpen
  \bibfield  {author} {\bibinfo {author} {\bibfnamefont {V.~M.}\ \bibnamefont
  {Acosta}}, \bibinfo {author} {\bibfnamefont {E.}~\bibnamefont {Bauch}},
  \bibinfo {author} {\bibfnamefont {A.}~\bibnamefont {Jarmola}}, \bibinfo
  {author} {\bibfnamefont {L.~J.}\ \bibnamefont {Zipp}}, \bibinfo {author}
  {\bibfnamefont {M.~P.}\ \bibnamefont {Ledbetter}}, \ and\ \bibinfo {author}
  {\bibfnamefont {D.}~\bibnamefont {Budker}},\ }\href {\doibase
  10.1063/1.3507884} {\bibfield  {journal} {\bibinfo  {journal} {Applied
  Physics Letters}\ }\textbf {\bibinfo {volume} {97}},\ \bibinfo {pages}
  {174104} (\bibinfo {year} {2010}{\natexlab{a}})}\BibitemShut {NoStop}%
\bibitem [{\citenamefont {Dolde}\ \emph {et~al.}(2011)\citenamefont {Dolde},
  \citenamefont {Fedder}, \citenamefont {Doherty}, \citenamefont {Nöbauer},
  \citenamefont {Rempp}, \citenamefont {Balasubramanian}, \citenamefont {Wolf},
  \citenamefont {Reinhard}, \citenamefont {Hollenberg}, \citenamefont
  {Jelezko},\ and\ \citenamefont {Wrachtrup}}]{dolde_electric-field_2011}%
  \BibitemOpen
  \bibfield  {author} {\bibinfo {author} {\bibfnamefont {F.}~\bibnamefont
  {Dolde}}, \bibinfo {author} {\bibfnamefont {H.}~\bibnamefont {Fedder}},
  \bibinfo {author} {\bibfnamefont {M.~W.}\ \bibnamefont {Doherty}}, \bibinfo
  {author} {\bibfnamefont {T.}~\bibnamefont {Nöbauer}}, \bibinfo {author}
  {\bibfnamefont {F.}~\bibnamefont {Rempp}}, \bibinfo {author} {\bibfnamefont
  {G.}~\bibnamefont {Balasubramanian}}, \bibinfo {author} {\bibfnamefont
  {T.}~\bibnamefont {Wolf}}, \bibinfo {author} {\bibfnamefont {F.}~\bibnamefont
  {Reinhard}}, \bibinfo {author} {\bibfnamefont {L.~C.~L.}\ \bibnamefont
  {Hollenberg}}, \bibinfo {author} {\bibfnamefont {F.}~\bibnamefont {Jelezko}},
  \ and\ \bibinfo {author} {\bibfnamefont {J.}~\bibnamefont {Wrachtrup}},\
  }\href {\doibase 10.1038/nphys1969} {\bibfield  {journal} {\bibinfo
  {journal} {Nature Physics}\ }\textbf {\bibinfo {volume} {7}},\ \bibinfo
  {pages} {459} (\bibinfo {year} {2011})}\BibitemShut {NoStop}%
\bibitem [{\citenamefont {Taylor}\ \emph {et~al.}(2008)\citenamefont {Taylor},
  \citenamefont {Cappellaro}, \citenamefont {Childress}, \citenamefont {Jiang},
  \citenamefont {Budker}, \citenamefont {Hemmer}, \citenamefont {Yacoby},
  \citenamefont {Walsworth},\ and\ \citenamefont
  {Lukin}}]{taylor_high-sensitivity_2008}%
  \BibitemOpen
  \bibfield  {author} {\bibinfo {author} {\bibfnamefont {J.~M.}\ \bibnamefont
  {Taylor}}, \bibinfo {author} {\bibfnamefont {P.}~\bibnamefont {Cappellaro}},
  \bibinfo {author} {\bibfnamefont {L.}~\bibnamefont {Childress}}, \bibinfo
  {author} {\bibfnamefont {L.}~\bibnamefont {Jiang}}, \bibinfo {author}
  {\bibfnamefont {D.}~\bibnamefont {Budker}}, \bibinfo {author} {\bibfnamefont
  {P.~R.}\ \bibnamefont {Hemmer}}, \bibinfo {author} {\bibfnamefont
  {A.}~\bibnamefont {Yacoby}}, \bibinfo {author} {\bibfnamefont
  {R.}~\bibnamefont {Walsworth}}, \ and\ \bibinfo {author} {\bibfnamefont
  {M.~D.}\ \bibnamefont {Lukin}},\ }\href {\doibase 10.1038/nphys1075}
  {\bibfield  {journal} {\bibinfo  {journal} {Nature Physics}\ }\textbf
  {\bibinfo {volume} {4}},\ \bibinfo {pages} {810} (\bibinfo {year}
  {2008})}\BibitemShut {NoStop}%
\bibitem [{\citenamefont {Loretz}\ \emph {et~al.}(2014)\citenamefont {Loretz},
  \citenamefont {Pezzagna}, \citenamefont {Meijer},\ and\ \citenamefont
  {Degen}}]{loretz_nanoscale_2014}%
  \BibitemOpen
  \bibfield  {author} {\bibinfo {author} {\bibfnamefont {M.}~\bibnamefont
  {Loretz}}, \bibinfo {author} {\bibfnamefont {S.}~\bibnamefont {Pezzagna}},
  \bibinfo {author} {\bibfnamefont {J.}~\bibnamefont {Meijer}}, \ and\ \bibinfo
  {author} {\bibfnamefont {C.~L.}\ \bibnamefont {Degen}},\ }\href {\doibase
  10.1063/1.4862749} {\bibfield  {journal} {\bibinfo  {journal} {Applied
  Physics Letters}\ }\textbf {\bibinfo {volume} {104}},\ \bibinfo {pages}
  {033102} (\bibinfo {year} {2014})}\BibitemShut {NoStop}%
\bibitem [{\citenamefont {Trusheim}\ and\ \citenamefont
  {Englund}(2016)}]{trusheim_wide-field_2016}%
  \BibitemOpen
  \bibfield  {author} {\bibinfo {author} {\bibfnamefont {M.~E.}\ \bibnamefont
  {Trusheim}}\ and\ \bibinfo {author} {\bibfnamefont {D.}~\bibnamefont
  {Englund}},\ }\href {\doibase 10.1088/1367-2630/aa5040} {\bibfield  {journal}
  {\bibinfo  {journal} {New Journal of Physics}\ }\textbf {\bibinfo {volume}
  {18}},\ \bibinfo {pages} {123023} (\bibinfo {year} {2016})}\BibitemShut
  {NoStop}%
\bibitem [{\citenamefont {Wolf}\ \emph {et~al.}(2015)\citenamefont {Wolf},
  \citenamefont {Rosenberg}, \citenamefont {Rapaport},\ and\ \citenamefont
  {Bar-Gill}}]{wolf_purcell-enhanced_2015}%
  \BibitemOpen
  \bibfield  {author} {\bibinfo {author} {\bibfnamefont {S.~A.}\ \bibnamefont
  {Wolf}}, \bibinfo {author} {\bibfnamefont {I.}~\bibnamefont {Rosenberg}},
  \bibinfo {author} {\bibfnamefont {R.}~\bibnamefont {Rapaport}}, \ and\
  \bibinfo {author} {\bibfnamefont {N.}~\bibnamefont {Bar-Gill}},\ }\href
  {\doibase 10.1103/PhysRevB.92.235410} {\bibfield  {journal} {\bibinfo
  {journal} {Physical Review B}\ }\textbf {\bibinfo {volume} {92}} (\bibinfo
  {year} {2015}),\ 10.1103/PhysRevB.92.235410}\BibitemShut {NoStop}%
\bibitem [{\citenamefont {Shields}\ \emph {et~al.}(2015)\citenamefont
  {Shields}, \citenamefont {Unterreithmeier}, \citenamefont {de~Leon},
  \citenamefont {Park},\ and\ \citenamefont {Lukin}}]{shields_efficient_2015}%
  \BibitemOpen
  \bibfield  {author} {\bibinfo {author} {\bibfnamefont {B.}~\bibnamefont
  {Shields}}, \bibinfo {author} {\bibfnamefont {Q.}~\bibnamefont
  {Unterreithmeier}}, \bibinfo {author} {\bibfnamefont {N.}~\bibnamefont
  {de~Leon}}, \bibinfo {author} {\bibfnamefont {H.}~\bibnamefont {Park}}, \
  and\ \bibinfo {author} {\bibfnamefont {M.}~\bibnamefont {Lukin}},\ }\href
  {\doibase 10.1103/PhysRevLett.114.136402} {\bibfield  {journal} {\bibinfo
  {journal} {Physical Review Letters}\ }\textbf {\bibinfo {volume} {114}}
  (\bibinfo {year} {2015}),\ 10.1103/PhysRevLett.114.136402}\BibitemShut
  {NoStop}%
\bibitem [{\citenamefont {Robledo}\ \emph {et~al.}(2011)\citenamefont
  {Robledo}, \citenamefont {Childress}, \citenamefont {Bernien}, \citenamefont
  {Hensen}, \citenamefont {Alkemade},\ and\ \citenamefont
  {Hanson}}]{robledo_high-fidelity_2011}%
  \BibitemOpen
  \bibfield  {author} {\bibinfo {author} {\bibfnamefont {L.}~\bibnamefont
  {Robledo}}, \bibinfo {author} {\bibfnamefont {L.}~\bibnamefont {Childress}},
  \bibinfo {author} {\bibfnamefont {H.}~\bibnamefont {Bernien}}, \bibinfo
  {author} {\bibfnamefont {B.}~\bibnamefont {Hensen}}, \bibinfo {author}
  {\bibfnamefont {P.~F.~A.}\ \bibnamefont {Alkemade}}, \ and\ \bibinfo {author}
  {\bibfnamefont {R.}~\bibnamefont {Hanson}},\ }\href {\doibase
  10.1038/nature10401} {\bibfield  {journal} {\bibinfo  {journal} {Nature}\
  }\textbf {\bibinfo {volume} {477}},\ \bibinfo {pages} {574} (\bibinfo {year}
  {2011})}\BibitemShut {NoStop}%
\bibitem [{\citenamefont {Hopper}\ \emph {et~al.}(2016)\citenamefont {Hopper},
  \citenamefont {Grote}, \citenamefont {Exarhos},\ and\ \citenamefont
  {Bassett}}]{hopper_near-infrared-assisted_2016}%
  \BibitemOpen
  \bibfield  {author} {\bibinfo {author} {\bibfnamefont {D.~A.}\ \bibnamefont
  {Hopper}}, \bibinfo {author} {\bibfnamefont {R.~R.}\ \bibnamefont {Grote}},
  \bibinfo {author} {\bibfnamefont {A.~L.}\ \bibnamefont {Exarhos}}, \ and\
  \bibinfo {author} {\bibfnamefont {L.~C.}\ \bibnamefont {Bassett}},\ }\href
  {\doibase 10.1103/PhysRevB.94.241201} {\bibfield  {journal} {\bibinfo
  {journal} {Physical Review B}\ }\textbf {\bibinfo {volume} {94}} (\bibinfo
  {year} {2016}),\ 10.1103/PhysRevB.94.241201}\BibitemShut {NoStop}%
\bibitem [{\citenamefont {Meirzada}\ \emph {et~al.}(2018)\citenamefont
  {Meirzada}, \citenamefont {Hovav}, \citenamefont {Wolf},\ and\ \citenamefont
  {Bar-Gill}}]{meirzada_negative_2017}%
  \BibitemOpen
  \bibfield  {author} {\bibinfo {author} {\bibfnamefont {I.}~\bibnamefont
  {Meirzada}}, \bibinfo {author} {\bibfnamefont {Y.}~\bibnamefont {Hovav}},
  \bibinfo {author} {\bibfnamefont {S.~A.}\ \bibnamefont {Wolf}}, \ and\
  \bibinfo {author} {\bibfnamefont {N.}~\bibnamefont {Bar-Gill}},\ }\href
  {\doibase 10.1103/PhysRevB.98.245411} {\bibfield  {journal} {\bibinfo
  {journal} {Phys. Rev. B}\ }\textbf {\bibinfo {volume} {98}},\ \bibinfo
  {pages} {245411} (\bibinfo {year} {2018})}\BibitemShut {NoStop}%
\bibitem [{\citenamefont {Ulbricht}\ and\ \citenamefont
  {Loh}(2018)}]{Ulbricht_Excited_state_lifetime_2018}%
  \BibitemOpen
  \bibfield  {author} {\bibinfo {author} {\bibfnamefont {R.}~\bibnamefont
  {Ulbricht}}\ and\ \bibinfo {author} {\bibfnamefont {Z.-H.}\ \bibnamefont
  {Loh}},\ }\href {\doibase 10.1103/PhysRevB.98.094309} {\bibfield  {journal}
  {\bibinfo  {journal} {Phys. Rev. B}\ }\textbf {\bibinfo {volume} {98}},\
  \bibinfo {pages} {094309} (\bibinfo {year} {2018})}\BibitemShut {NoStop}%
\bibitem [{\citenamefont {Acosta}\ \emph
  {et~al.}(2010{\natexlab{b}})\citenamefont {Acosta}, \citenamefont {Jarmola},
  \citenamefont {Bauch},\ and\ \citenamefont {Budker}}]{acosta_optical_2010}%
  \BibitemOpen
  \bibfield  {author} {\bibinfo {author} {\bibfnamefont {V.~M.}\ \bibnamefont
  {Acosta}}, \bibinfo {author} {\bibfnamefont {A.}~\bibnamefont {Jarmola}},
  \bibinfo {author} {\bibfnamefont {E.}~\bibnamefont {Bauch}}, \ and\ \bibinfo
  {author} {\bibfnamefont {D.}~\bibnamefont {Budker}},\ }\href {\doibase
  10.1103/PhysRevB.82.201202} {\bibfield  {journal} {\bibinfo  {journal}
  {Physical Review B}\ }\textbf {\bibinfo {volume} {82}} (\bibinfo {year}
  {2010}{\natexlab{b}}),\ 10.1103/PhysRevB.82.201202}\BibitemShut {NoStop}%
\bibitem [{\citenamefont {Livneh}\ \emph {et~al.}(2011)\citenamefont {Livneh},
  \citenamefont {Strauss}, \citenamefont {Schwarz}, \citenamefont {Rosenberg},
  \citenamefont {Zimran}, \citenamefont {Yochelis}, \citenamefont {Chen},
  \citenamefont {Banin}, \citenamefont {Paltiel},\ and\ \citenamefont
  {Rapaport}}]{livneh_highly_2011}%
  \BibitemOpen
  \bibfield  {author} {\bibinfo {author} {\bibfnamefont {N.}~\bibnamefont
  {Livneh}}, \bibinfo {author} {\bibfnamefont {A.}~\bibnamefont {Strauss}},
  \bibinfo {author} {\bibfnamefont {I.}~\bibnamefont {Schwarz}}, \bibinfo
  {author} {\bibfnamefont {I.}~\bibnamefont {Rosenberg}}, \bibinfo {author}
  {\bibfnamefont {A.}~\bibnamefont {Zimran}}, \bibinfo {author} {\bibfnamefont
  {S.}~\bibnamefont {Yochelis}}, \bibinfo {author} {\bibfnamefont
  {G.}~\bibnamefont {Chen}}, \bibinfo {author} {\bibfnamefont {U.}~\bibnamefont
  {Banin}}, \bibinfo {author} {\bibfnamefont {Y.}~\bibnamefont {Paltiel}}, \
  and\ \bibinfo {author} {\bibfnamefont {R.}~\bibnamefont {Rapaport}},\ }\href
  {\doibase 10.1021/nl200052j} {\bibfield  {journal} {\bibinfo  {journal} {Nano
  Letters}\ }\textbf {\bibinfo {volume} {11}},\ \bibinfo {pages} {1630}
  (\bibinfo {year} {2011})}\BibitemShut {NoStop}%
\bibitem [{\citenamefont {Harats}\ \emph {et~al.}(2014)\citenamefont {Harats},
  \citenamefont {Livneh}, \citenamefont {Zaiats}, \citenamefont {Yochelis},
  \citenamefont {Paltiel}, \citenamefont {Lifshitz},\ and\ \citenamefont
  {Rapaport}}]{harats_full_2014}%
  \BibitemOpen
  \bibfield  {author} {\bibinfo {author} {\bibfnamefont {M.~G.}\ \bibnamefont
  {Harats}}, \bibinfo {author} {\bibfnamefont {N.}~\bibnamefont {Livneh}},
  \bibinfo {author} {\bibfnamefont {G.}~\bibnamefont {Zaiats}}, \bibinfo
  {author} {\bibfnamefont {S.}~\bibnamefont {Yochelis}}, \bibinfo {author}
  {\bibfnamefont {Y.}~\bibnamefont {Paltiel}}, \bibinfo {author} {\bibfnamefont
  {E.}~\bibnamefont {Lifshitz}}, \ and\ \bibinfo {author} {\bibfnamefont
  {R.}~\bibnamefont {Rapaport}},\ }\href {\doibase 10.1021/nl502652k}
  {\bibfield  {journal} {\bibinfo  {journal} {Nano Letters}\ }\textbf {\bibinfo
  {volume} {14}},\ \bibinfo {pages} {5766} (\bibinfo {year}
  {2014})}\BibitemShut {NoStop}%
\bibitem [{\citenamefont {Wan}\ \emph {et~al.}(2018)\citenamefont {Wan},
  \citenamefont {Mouradian},\ and\ \citenamefont
  {Englund}}]{wan_two-dimensional_2018}%
  \BibitemOpen
  \bibfield  {author} {\bibinfo {author} {\bibfnamefont {N.~H.}\ \bibnamefont
  {Wan}}, \bibinfo {author} {\bibfnamefont {S.}~\bibnamefont {Mouradian}}, \
  and\ \bibinfo {author} {\bibfnamefont {D.}~\bibnamefont {Englund}},\ }\href
  {\doibase 10.1063/1.5021349} {\bibfield  {journal} {\bibinfo  {journal}
  {Applied Physics Letters}\ }\textbf {\bibinfo {volume} {112}},\ \bibinfo
  {pages} {141102} (\bibinfo {year} {2018})}\BibitemShut {NoStop}%
\bibitem [{\citenamefont {Mouradian}\ \emph {et~al.}(2017)\citenamefont
  {Mouradian}, \citenamefont {Wan}, \citenamefont {Schröder},\ and\
  \citenamefont {Englund}}]{mouradian_rectangular_2017}%
  \BibitemOpen
  \bibfield  {author} {\bibinfo {author} {\bibfnamefont {S.}~\bibnamefont
  {Mouradian}}, \bibinfo {author} {\bibfnamefont {N.~H.}\ \bibnamefont {Wan}},
  \bibinfo {author} {\bibfnamefont {T.}~\bibnamefont {Schröder}}, \ and\
  \bibinfo {author} {\bibfnamefont {D.}~\bibnamefont {Englund}},\ }\href
  {\doibase 10.1063/1.4992118} {\bibfield  {journal} {\bibinfo  {journal}
  {Applied Physics Letters}\ }\textbf {\bibinfo {volume} {111}},\ \bibinfo
  {pages} {021103} (\bibinfo {year} {2017})}\BibitemShut {NoStop}%
\bibitem [{\citenamefont {Haroche}\ and\ \citenamefont
  {Kleppner}(1989)}]{haroche_cavity_1989}%
  \BibitemOpen
  \bibfield  {author} {\bibinfo {author} {\bibfnamefont {S.}~\bibnamefont
  {Haroche}}\ and\ \bibinfo {author} {\bibfnamefont {D.}~\bibnamefont
  {Kleppner}},\ }\href {\doibase 10.1063/1.881201} {\bibfield  {journal}
  {\bibinfo  {journal} {Physics Today}\ }\textbf {\bibinfo {volume} {42}},\
  \bibinfo {pages} {24} (\bibinfo {year} {1989})}\BibitemShut {NoStop}%
\bibitem [{\citenamefont {Minkov}\ and\ \citenamefont
  {Savona}(2014)}]{minkov_automated_2014}%
  \BibitemOpen
  \bibfield  {author} {\bibinfo {author} {\bibfnamefont {M.}~\bibnamefont
  {Minkov}}\ and\ \bibinfo {author} {\bibfnamefont {V.}~\bibnamefont
  {Savona}},\ }\href {https://doi.org/10.1038/srep05124} {\bibfield  {journal}
  {\bibinfo  {journal} {Scientific Reports}\ }\textbf {\bibinfo {volume} {4}},\
  \bibinfo {pages} {5124} (\bibinfo {year} {2014})}\BibitemShut {NoStop}%
\bibitem [{\citenamefont {Zhang}\ \emph {et~al.}(2018)\citenamefont {Zhang},
  \citenamefont {Sun}, \citenamefont {Burek}, \citenamefont {Dory},
  \citenamefont {Tzeng}, \citenamefont {Fischer}, \citenamefont {Kelaita},
  \citenamefont {Lagoudakis}, \citenamefont {Radulaski}, \citenamefont {Shen},
  \citenamefont {Melosh}, \citenamefont {Chu}, \citenamefont {Lončar},\ and\
  \citenamefont {Vučković}}]{zhang_strongly_2018}%
  \BibitemOpen
  \bibfield  {author} {\bibinfo {author} {\bibfnamefont {J.~L.}\ \bibnamefont
  {Zhang}}, \bibinfo {author} {\bibfnamefont {S.}~\bibnamefont {Sun}}, \bibinfo
  {author} {\bibfnamefont {M.~J.}\ \bibnamefont {Burek}}, \bibinfo {author}
  {\bibfnamefont {C.}~\bibnamefont {Dory}}, \bibinfo {author} {\bibfnamefont
  {Y.-K.}\ \bibnamefont {Tzeng}}, \bibinfo {author} {\bibfnamefont {K.~A.}\
  \bibnamefont {Fischer}}, \bibinfo {author} {\bibfnamefont {Y.}~\bibnamefont
  {Kelaita}}, \bibinfo {author} {\bibfnamefont {K.~G.}\ \bibnamefont
  {Lagoudakis}}, \bibinfo {author} {\bibfnamefont {M.}~\bibnamefont
  {Radulaski}}, \bibinfo {author} {\bibfnamefont {Z.-X.}\ \bibnamefont {Shen}},
  \bibinfo {author} {\bibfnamefont {N.~A.}\ \bibnamefont {Melosh}}, \bibinfo
  {author} {\bibfnamefont {S.}~\bibnamefont {Chu}}, \bibinfo {author}
  {\bibfnamefont {M.}~\bibnamefont {Lončar}}, \ and\ \bibinfo {author}
  {\bibfnamefont {J.}~\bibnamefont {Vučković}},\ }\href {\doibase
  10.1021/acs.nanolett.7b05075} {\bibfield  {journal} {\bibinfo  {journal}
  {Nano Letters}\ }\textbf {\bibinfo {volume} {18}},\ \bibinfo {pages} {1360}
  (\bibinfo {year} {2018})}\BibitemShut {NoStop}%
\bibitem [{\citenamefont {Pham}(2013)}]{Linh_thesis}%
  \BibitemOpen
  \bibfield  {author} {\bibinfo {author} {\bibfnamefont {L.}~\bibnamefont
  {Pham}},\ }\emph {\bibinfo {title} {Magnetic Field Sensing with
  Nitrogen-Vacancy Color Centers in Diamond.}},\ \href@noop {} {Ph.D. thesis},\
  \bibinfo  {school} {Harvard University} (\bibinfo {year} {2013})\BibitemShut
  {NoStop}%
\bibitem [{\citenamefont {Kehayias}\ \emph {et~al.}(2013)\citenamefont
  {Kehayias}, \citenamefont {Doherty}, \citenamefont {English}, \citenamefont
  {Fischer}, \citenamefont {Jarmola}, \citenamefont {Jensen}, \citenamefont
  {Leefer}, \citenamefont {Hemmer}, \citenamefont {Manson},\ and\ \citenamefont
  {Budker}}]{PhysRevB.88.165202}%
  \BibitemOpen
  \bibfield  {author} {\bibinfo {author} {\bibfnamefont {P.}~\bibnamefont
  {Kehayias}}, \bibinfo {author} {\bibfnamefont {M.~W.}\ \bibnamefont
  {Doherty}}, \bibinfo {author} {\bibfnamefont {D.}~\bibnamefont {English}},
  \bibinfo {author} {\bibfnamefont {R.}~\bibnamefont {Fischer}}, \bibinfo
  {author} {\bibfnamefont {A.}~\bibnamefont {Jarmola}}, \bibinfo {author}
  {\bibfnamefont {K.}~\bibnamefont {Jensen}}, \bibinfo {author} {\bibfnamefont
  {N.}~\bibnamefont {Leefer}}, \bibinfo {author} {\bibfnamefont
  {P.}~\bibnamefont {Hemmer}}, \bibinfo {author} {\bibfnamefont {N.~B.}\
  \bibnamefont {Manson}}, \ and\ \bibinfo {author} {\bibfnamefont
  {D.}~\bibnamefont {Budker}},\ }\href {\doibase 10.1103/PhysRevB.88.165202}
  {\bibfield  {journal} {\bibinfo  {journal} {Phys. Rev. B}\ }\textbf {\bibinfo
  {volume} {88}},\ \bibinfo {pages} {165202} (\bibinfo {year}
  {2013})}\BibitemShut {NoStop}%
\bibitem [{\citenamefont {Dumeige}\ \emph {et~al.}(2013)\citenamefont
  {Dumeige}, \citenamefont {Chipaux}, \citenamefont {Jacques}, \citenamefont
  {Treussart}, \citenamefont {Roch}, \citenamefont {Debuisschert},
  \citenamefont {Acosta}, \citenamefont {Jarmola}, \citenamefont {Jensen},
  \citenamefont {Kehayias},\ and\ \citenamefont
  {Budker}}]{dumeige_magnetometry_2013}%
  \BibitemOpen
  \bibfield  {author} {\bibinfo {author} {\bibfnamefont {Y.}~\bibnamefont
  {Dumeige}}, \bibinfo {author} {\bibfnamefont {M.}~\bibnamefont {Chipaux}},
  \bibinfo {author} {\bibfnamefont {V.}~\bibnamefont {Jacques}}, \bibinfo
  {author} {\bibfnamefont {F.}~\bibnamefont {Treussart}}, \bibinfo {author}
  {\bibfnamefont {J.-F.}\ \bibnamefont {Roch}}, \bibinfo {author}
  {\bibfnamefont {T.}~\bibnamefont {Debuisschert}}, \bibinfo {author}
  {\bibfnamefont {V.~M.}\ \bibnamefont {Acosta}}, \bibinfo {author}
  {\bibfnamefont {A.}~\bibnamefont {Jarmola}}, \bibinfo {author} {\bibfnamefont
  {K.}~\bibnamefont {Jensen}}, \bibinfo {author} {\bibfnamefont
  {P.}~\bibnamefont {Kehayias}}, \ and\ \bibinfo {author} {\bibfnamefont
  {D.}~\bibnamefont {Budker}},\ }\href {\doibase 10.1103/PhysRevB.87.155202}
  {\bibfield  {journal} {\bibinfo  {journal} {Physical Review B}\ }\textbf
  {\bibinfo {volume} {87}} (\bibinfo {year} {2013}),\
  10.1103/PhysRevB.87.155202}\BibitemShut {NoStop}%
\end{thebibliography}%

\end{document}